\documentclass{ecai}
\usepackage{times}
\usepackage{graphicx}
\usepackage{latexsym}
\usepackage{amsmath}
\usepackage{booktabs}
\usepackage{amssymb}
\usepackage{hyperref}

\usepackage{caption}
\usepackage{subcaption}

\begin{document}

\title{Learning Style-Aware Symbolic Music Representations by Adversarial Autoencoders}

\author{AndreaValenti\institute{Corresponding author: andrea.valenti@phd.unipi.it}  \footnotemark[2] \and Antonio Carta\footnotemark[2] \and Davide Bacciu\institute{Dipartimento di Informatica, Universit\`{a} di Pisa, Italy. \newline Emails: \{antonio.carta, davide.bacciu\}@di.unipi.it, \newline andrea.valenti@phd.unipi.it} 
}

\maketitle

\begin{abstract}
We address the challenging open problem of learning an effective latent space for symbolic music data in generative music modeling. We focus on leveraging adversarial regularization as a flexible and natural mean to imbue variational autoencoders with context information concerning music genre and style. Through the paper, we show how Gaussian mixtures taking into account music metadata information can be used as an effective prior for the autoencoder latent space, introducing the first Music Adversarial Autoencoder (MusAE). The empirical analysis on a large scale benchmark shows that our model has a higher reconstruction accuracy than state-of-the-art models based on standard variational autoencoders. It is also able to create realistic interpolations between two musical sequences, smoothly changing the dynamics of the different tracks. Experiments show that the model can organise its latent space accordingly to low-level properties of the musical pieces, as well as to embed into the latent variables the high-level genre information injected from the prior distribution to increase its overall performance. This allows us to perform changes to the generated pieces in a principled way.
\end{abstract}

\section{Introduction}
In recent years, we have been witnessing an increasing attention on \emph{creative aspects} of artificial intelligence, in particular when linked to the generation of art, such as in painting\cite{li2017demystifying_style_transfer,jing2019style_transfer_review,gatys2016image_style_transfer}, literature\cite{hu2017controlled_text_generation,shen2017_parallel_style_text} and music composition\cite{RobertsEngelRaffel, hadjeres2016deepbach}. Much of this interest has been enabled by the availability of generative deep learning models apt to model distributions or sampling processes over such intricate and multifaceted pieces of information (begging pardon for the simplification of such a high act of human creativity which is art). While these models appear sufficiently mature to tackle the generation of plausible artificial works in some artistic fields, such as in visual arts, there are other arts in which we are still struggling to capture the essence and variability of the underlying stylistic aspects. Music is one of such arts, characterized by still inextricable entanglements between different stylistic dimensions.

In this paper, we propose a model whose long-term ambition can be that to assist a composer during the writing process of a novel score. The model is composed of two parts: an encoder and a decoder. The encoder takes as input a score, represented as several tracks for each instrument, and computes a latent vector. Ideally, the latent vector should encode the entire score into musically meaningful features, easy to understand from a musician's perspective and disentangled such that each property of the original score can be customized separately with a simple modification to the vector. A composer can modify the vector to change low-level properties of the score, such as the note's density or average length, or it can modify high-level features like the genre, for example, to create a more rock or jazzy tune. The modified vector can be fed as input to the decoder, which generates a novel score with the desired features.
This approach requires to solve a fundamental question: \emph{How can we learn a good representation for a music score?}

In this paper, we present a novel architecture for music modeling and generation. Its name is MusAE (the name is a reference to the Muses of ancient Greek mythology, the inspirational goddesses, of Literature, Science and Art) and its architecture is based on the adversarial autoencoder (AAE) \cite{Makhzani}.
AAEs give us several advantages over standard variational autoencoders (VAEs): using adversarial regularization in place of the KL divergence allows for more flexibility over the choice of prior distribution of the latent variables, which can, in principle, be forced to follow any probability distribution, even ones with unknown functional form (for example, in one of their experiments Makhzani et al. trained an AAE whose latent factors are forced to follow a swiss roll distribution \cite{Makhzani}).
This constitutes an important advantage for AAEs over VAEs, where the imposed prior distribution must necessarily be chosen to make the KL divergence term computable. MusAE is capable of reconstructing musical phrases with high accuracy and to interpolate between latent representations in a sound pleasing way. The model can also change specific properties of the songs by modifying their respective latent factors. To the best of our knowledge, this is the first time an AAE has been applied with success to automatic music generation.

\section{Background}
\subsection{Related Work}
As surveyed in \cite{Briot}, recent years have seen an increasingly successful application of generative models such as the VAE by \cite{KingmaWelling} and generative adversarial networks (GAN) by \cite{Goodfellow}. The MIDI-VAE model by \cite{Brunner} is based on VAEs and it considers both pitches, velocities and instruments to perform style transfer between two genres. 
In \cite{RobertsEngelRaffel}, it is introduced Music VAE, a hierarchical sequence-to-sequence VAE model. The hierarchy is expressed by a two-level decoder: in the first level the so-called ``conductor" decoder takes the latent factors generated by the encoder as input, and in turn generates some intermediates codes, which are then used by the low-level decoder to generate the actual piano roll. The hierarchical structure of Music VAE makes it able to capture long-term structure in polyphonic music, which is generally considered to be a difficult problem, due to the complexity of the data distribution.

GAN, albeit powerful in principle, are very difficult to train effectively \cite{ArjovskyBottou}, especially when applied to sequential data. However, this has not discouraged scientists, and many recent works successfully apply GAN to music generation. Works by \cite{Mogren}, \cite{Yang} and \cite{Dong} have shown that different types of GAN can be effectively applied to music composition. Finally, the SeqGAN model \cite{Yu}, applies an RNN-based GAN to music by incorporating reinforcement learning techniques.

\subsection{Adversarial Autoencoder}
First introduced by \cite{Makhzani}, the AAE model can be viewed as the combination of VAE and GAN.
Formally, let $x$ be the data sample and $z$ be the corresponding latent variables of the AAE. Let $p(z)$ be the prior imposed on the latent variables, let $q_\lambda(z|x)$ be the encoder conditional distribution and let $p_\theta(x|z)$ be the decoder conditional distribution. Let $p_d(x)$ the actual data distribution and $p_g(x)$ be the model distribution (i.e. the distribution of data generated by the model). Then, the encoder defines an aggregated posterior distribution $q_\lambda(z)$ on the latent variables $z$ as follows
\begin{equation}
    q_\lambda(z) = \int q_\lambda(z|x)p_d(x)
\end{equation}

In AAEs the aggregate posterior $q_\lambda(z)$ is made to match the arbitrary prior $p(z)$ by attaching an adversarial network (the discriminator) at the top of the latent variables $z$. The loss function of AAE thus becomes
\begin{equation}
    \label{eq_aae_loss}
    \mathcal{L}_{AAE} = \mathbb{E}_{z \sim q_\lambda(z|x)} \lbrack log p_\theta(x|z) \rbrack + \beta \mathcal{L}_{GAN}
\end{equation}
where $\beta$ regulates the amount of desired adversarial regularisation. $\mathcal{L}_{GAN}$ is the adversarial GAN loss as defined in \cite{Goodfellow}.

During training, the autoencoder learns to minimise the reconstruction error, while the adversarial network guides the encoder to match the imposed prior. Thus, the encoder plays the role of the generator of the adversarial part, while the adversarial network is equivalent to the GAN's discriminator. At the end of the training, the decoder can be used as a generative model that maps the imposed prior $p(z)$ to the data distribution.

The training procedure of an AAE consists of two phases: the reconstruction phase and the regularization phase, corresponding to a loss $\mathcal{L}_{total}$  comprising two terms
\begin{equation}
    \label{eq_musae_loss}
    \mathcal{L}_{total} = \mathcal{H}(p_{d}, p_{g}) + \beta \mathcal{L}_{WGAN-GP},
\end{equation}
where $\mathcal{H}(p_{d}, p_{g})$ denotes the cross entropy between the original data samples and their reconstructions, and acts as autoencoder loss during the reconstruction phase.
The second term is the Wasserstein GAN with Gradient Penalty (WGAN-GP)  loss, \cite{Gulrajani} that is
\begin{align}
\label{eq_wgan-gp}
    \mathcal{L}_{WGAN-GP} &= \mathbb{E}_{x \sim p_r} \lbrack d_\phi(x) \rbrack\\
     & - \mathbb{E}_{z \sim p_z}[d_\phi(g_\theta(z))]\\
      & + \lambda \mathbb{E}_{\hat{x} \sim p_{\hat{x}}}
      [(\Vert \nabla_{\hat{x}} d_\phi(\hat{x}) \Vert_{2} - 1)^{2}]
\end{align}
where $\hat{x}$ is computed by $\hat{x} = \alpha g_\theta(z) + (1-\alpha) x$ and $\alpha$ is sampled uniformly between 0 and 1.
This term is responsible for the adversarial regularization imposed on the latent variables $z$. Each phase is trained with gradient descent end-to-end.

\section{Music Adversarial Autoencoder (MusAE)}
MusAE is based on an AAE architecture, exploiting the adversarial regularization to allow richer representations of the latent space instead of constraining it to a simple Gaussian.
This section describes the model architecture and the input representation. 

\subsection{Data Representation}
\label{sec_datarep}
The model is designed to work with multitrack MIDI songs. During the preprocessing stage, the raw MIDI format is transformed into a multitrack piano roll representation. Each song becomes a sequence, sampled at regular intervals, where each timestep contains the information to play each track. We use a resolution of 4 timesteps per quarter note, for a total of 16 timesteps per bar. For a given track, each timestep reports a categorical variable, taking one of the 130 binary discrete states:
\begin{itemize}
\item 128 states representing the possible values for MIDI pitches\cite{MIDI}.
\item One ``hold note" state, which indicates to keep playing the note which was played at the previous timestep.
\item One ``silent note" state, which indicates a timestep when no notes are playing.
\end{itemize}
In the following experiments, we consider four-track songs as this is the information available in the Lakh Dataset used in the experimental analysis (see Section \ref{sec_dataset}). 
Each sample is thus represented by a $n_{timesteps} \times n_{notes} \times n_{tracks}$ matrix.
Each song in the dataset has metadata information associated with it, such as title, author, and so on. In particular, we consider the genre tags of songs. These tags will be useful for ``inspiring" the choice of a prior distribution of the latent variables (for more information see Sect. \ref{rec_exp_setup})

\subsection{Architecture}
\begin{figure}
    \centering
    \includegraphics[width=\linewidth]{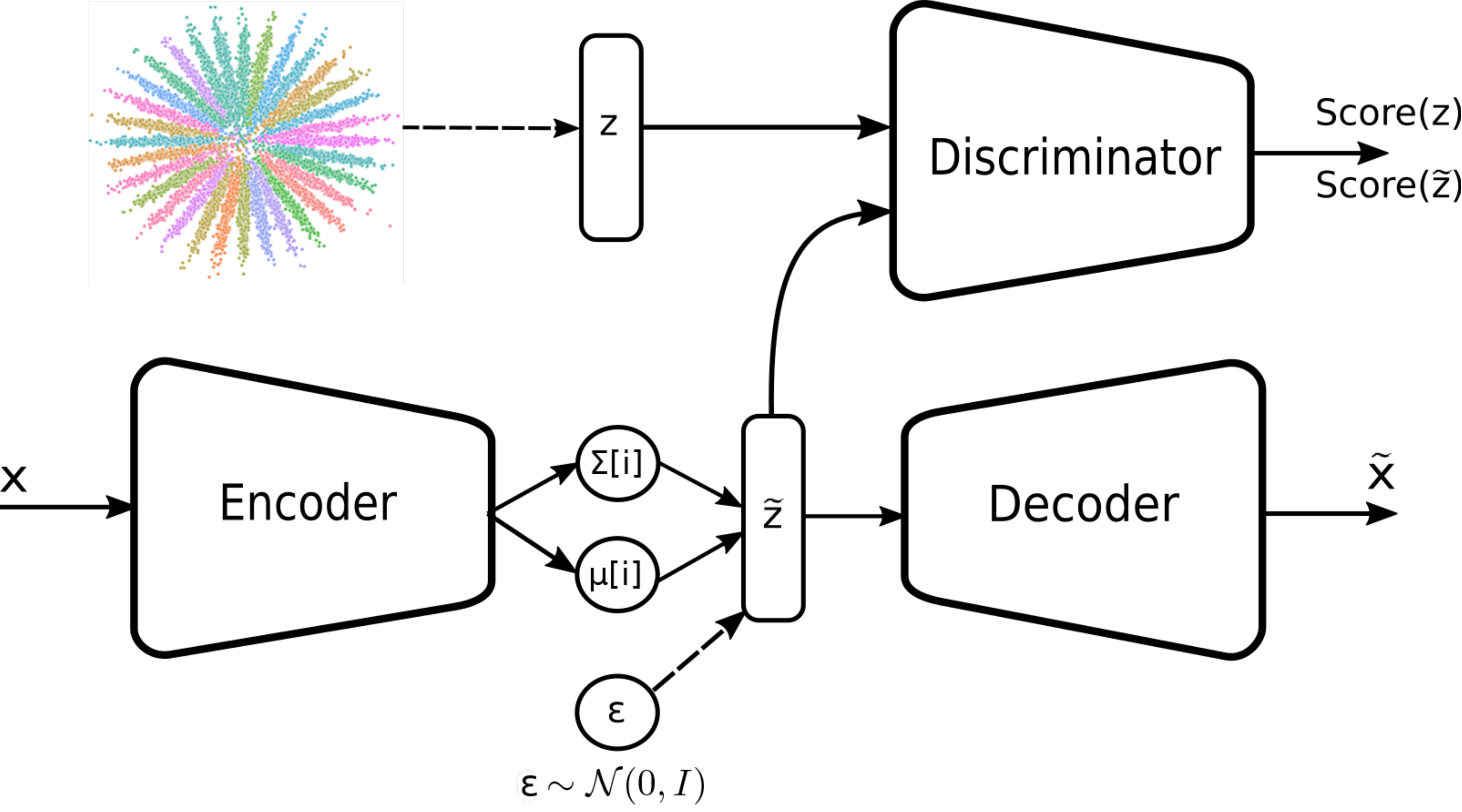}
    \caption{MusAE's architecture. Dashed lines denote stochastic operations, full lines denotes deterministic operations.}
    \label{fig:aae}
\end{figure}

MusAE is based on an AAE architecture, comprising an autoencoding part responsible for song reconstruction and an adversarial part injecting a regularization penalty \cite{Makhzani}. In this section, we describe the main components of the architecture (depicted in Figure \ref{fig:aae}): the encoder, the decoder and the discriminator. In general, the architecture and hyperparameters have been chosen to be as close as possible to \cite{RobertsEngelRaffel}, in order to perform a direct comparison of the empirical results.

\subsubsection{Encoder}

The encoder takes the multitrack piano roll representation and computes the latent factors $z \in \mathbb{R}^{N_z}$ that encode the entire sequence. $N_z = 512$ is the size of the latent space. The architecture, shown in Figure \ref{fig:aae}, is composed of $3$ layers of bidirectional LSTMs (biLSTM \cite{Schuster}) with $N_h = 1024$ hidden units. The bidirectionality of the model enables the encoder to produce more informative latent variables by using context information about future timesteps: this is a desirable property in the case of music data since the meaning of a note usually depends on the whole phrase and it is not only restricted to the preceding notes. 
Let $x$ be a training sample. Then, the encoder's behaviour can be described by
\begin{align}
    h =& biLSTM(x) \\
    \mu =& W_{\mu} h \\
    \sigma_{pre} =& W_{\sigma} h \\
    \sigma =& exp(2\sigma_{pre}),
\end{align}
where $W_{\mu} \in \mathbb{R}^{N_h \times N_z}$ and $W_{\sigma} \in \mathbb{R}^{N_h \times N_z}$ are the weight matrices of the linear layers used to compute, respectively, $\mu$ and $\sigma_{pre}$.


\subsubsection{Decoder}
The decoder performs the inverse operation of the encoder: given the latent factors $z$, it produces the piano roll corresponding to that particular combination of latent variables (see Figure \ref{fig:aae}). We use a separate decoder for each track, allowing each module to specialize in the different dynamics of the tracks.
Let $z$ be a latent code, the decoder operates as follows:
\begin{align}
    h^{track}_0 &= W_{track} z\\
    \tilde{x}_{track} &= Softmax(LSTM_{track}(z, h^{track}_0)) \\
    \tilde{x} &= [\tilde{x}_{drums} | \tilde{x}_{bass} | \tilde{x}_{guitar} | \tilde{x}_{strings}],
\end{align}
where $track \in \{drums, bass, guitar, strings\}$, $W_{track} \in \mathbb{R}^{N_z \times N_h}$ denotes the weight parameters of the linear layers used to compute the initial hidden state $h^{track}_0 \in \mathbb{R}^{N_h}$ for their respective LSTMs, $\tilde{x}$ is the track generated from the decoder, and the symbol $|$ denotes the concatenation operator. Each decoder comprises 3-layers of LSTMs with $N_h$ hidden units. The latent vector $z$ is fed as a constant input to the LSTMs at each timestep.


In contrast with other related models (\cite{RobertsEngelRaffel} \cite{Brunner}), our decoder is not autoregressive, meaning that the only information available during the song generation is the latent code $z$. This limitation constrains the decoder to exploit the latent code to generate the sequence and eliminates the tendency of ignoring the latent code \cite{RobertsEngelRaffel} that is typical of powerful autoregressive models.

\subsubsection{Discriminator}
The discriminator is a multilayer perceptron with $2$ hidden layers of $N_h$ hidden units each followed by a \textit{tanh} activation. Its task is to distinguish between ``real" latent variables, sampled from prior latent distributions, and ``artificial" ones, generated by the encoder to perform the variational approximation of the ELBO in an adversarial way. 
Following the literature for the Wasserstein GAN \cite{Arjovsky}, the last layer of the discriminator is a linear layer which outputs the final score, that is, a scalar measure of how well a latent variable resembles the imposed prior distribution.

\subsection{Choice of Prior Distributions}
\label{sec_choice_of_prior}
Using an adversarial discriminator network in place of the KL divergence to perform latent space regularisation allows us more flexibility in the choice of the prior distribution of the latent factors $z$. While, for standard VAE, we are required to use a distribution for which a closed-form of the KL divergence is known and tractable (typically, in practice, this limits the choice to gaussian distributions), by using an AAE the only requirement we have is to be able to sample from the chosen distribution. This allows us to choose non-standard distributions that may, for example, leverage any prior knowledge that is specific from the particular domain of application.
In particular, for the experiments described in the next sections, we decided to impose two different prior distribution. First, as a baseline, we used a standard isotropic gaussian with mean $0$ and variance $I$, much like VAEs. This allowed us to perform a direct comparison of our results with other VAE models found in the literature. The second distribution is obtained from a mixture of 32 gaussians, displaced in a ``flower-like" shape (it can be thought as a $N_z$-dimensional generalization of the distribution depicted in Figure 4a of \cite{Makhzani}). This ``flower distribution" allows us to put similar genres, which are going to have similar metrics (see Fig. \ref{fig:relative_changes}), closer in the latent space. This is a beneficial constraint that is not easy to enforce by using only a single gaussian. Furthermore, it allows to model ``cross-genres" songs by pushing their latent codes to the center of the ``flower". Formally, the means of such gaussians are displaced on a 2-dimensional circle, according to the formula
\begin{equation}
    \mu_i = [cos(\frac{2\pi i}{n}), sin(\frac{2\pi i}{n}), 0, ..., 0]
\end{equation}
where $n=32$ in our case. The $\mu_i$ have a total of $N_z$ dimensions. The covariance matrix $\Sigma_i$ is obtained in the following way: 
\begin{align}
    v_i &= \begin{bmatrix} 
        cos(\frac{2\pi i}{n}) & sin(\frac{2\pi i}{n}) \\
        -sin(\frac{2\pi i}{n}) & cos(\frac{2\pi i}{n})
    \end{bmatrix} \\
    M_i &= \begin{bmatrix} 
        v_i^T & 0 \\
        0 & I
    \end{bmatrix} \\
    S &= \begin{bmatrix} 
        a_1 & 0 \\
        0 & diag(a_2)
    \end{bmatrix}\\
    \Sigma_i &= M_iSM_i^{-1}
\end{align}
where, as before, $n=32$ and $M_i$ and $S$ are $N_z\times N_z$ matrices. 
In practice, we set variance $a_1=0.1$ for the radial (i.e. center-to-outer) dimension, and variance $a_2=0.001$ for the remaining dimensions. We then use the matrix $M_i$ to rotate $S$ accordingly to the position of the specific gaussian.
We chose to keep just the 32 most-occurring genre tags to avoid the noisier ones. Each genre tag is associated with a different gaussian in the latent space, based on the metadata information that comes with each song (see Sect. \ref{sec_datarep}). The gaussian for each genre was chosen in order to put related genres closer in the latent space, by assigning them to gaussians that are closer to each other. By injecting genre information we are thus able to encourage a meaningful organisation of the latent space. 
If a song had more than one tag associated with it, we randomly selected a single tag using uniform probability (this selection happens at training time, so the training process can select another available tag for that song in a subsequent epoch).

\section{Experimental Analysis}

\begin{table*}[t]
    \centering
    \caption{Reconstruction accuracy of MusAE on the Lakh MIDI dataset, compared to the Music VAE model\protect\cite{RobertsEngelRaffel}. Higher is better.}
    \begin{tabular}{lcccc}
    \toprule
    &   MusAE (gaussian)    &  MusAE (gaussian mixture) & Music VAE (flat)  & Music VAE (hierarchical) \\ \midrule
    2 bar drums         &   0.999     &   --         &       0.917      &   -- \\
    2-bar melody        &   0.991     &   --         &       0.951      &   -- \\ \midrule
    
    16-bar drums        &   0.773     &   0.911         &       0.641      &   0.895 \\
    16-bar bass         &   0.763     &   0.880         &       0.651      &   0.782 \\
    16-bar guitar/piano &   0.710     &   0.797         &       0.660      &   0.760 \\
    16-bar strings      &   0.802     &   0.825         &       --         &    --   \\
    16-bar average (dbgs)      &   0.762     &   0.853         &       --      &   -- \\
    16-bar average (dbg)      &   0.749     &   0.863         &       0.651      &   0.812  \\
    \bottomrule
    \end{tabular}
    \label{tab_accuracy}
\end{table*}

In this section, we empirically assess the ability of MusAE to properly encode musical sequences and to learn a structured and interpretable organization of the latent space.

A useful music editor must have several desirable properties. 

First, it should be able to represent different input songs with high accuracy. This is especially important for music, where even little variations can significantly alter the structure and the final appeal of a song. Specifically for our model, we must test the ability of the learned latent space to be expressive enough to capture every subtle nuance of the input data.

Then, the model should be able to capture several properties of music. Ideally, the model should separate the content from the style, and allow the user to easily modify each property of the final song. Therefore, the learned latent space of the model should be able to effectively capture the important factors of variation of the data space. Given two different songs mapped into the latent space, the model must allow to easily sample new points from the latent space (e.g. with linear interpolation) and get corresponding points in the data space that still sound realistic, and whose musical properties vary smoothly. Indeed, identifying the meaningful axis of variation in the latent space is the first step for being able to modify a song in a principled and controllable way.

For reproducibility, and to facilitate future research in the field, we make the code publicly available on Github \footnote{\url{https://github.com/Andrea-V/MusAE}}. 
It is also possible to listen to some of the final generated samples in the additional material \footnote{\url{http://bit.ly/MusAE-ADD}}.

\begin{figure*}[t]
  \centering
  \includegraphics[width=\linewidth]{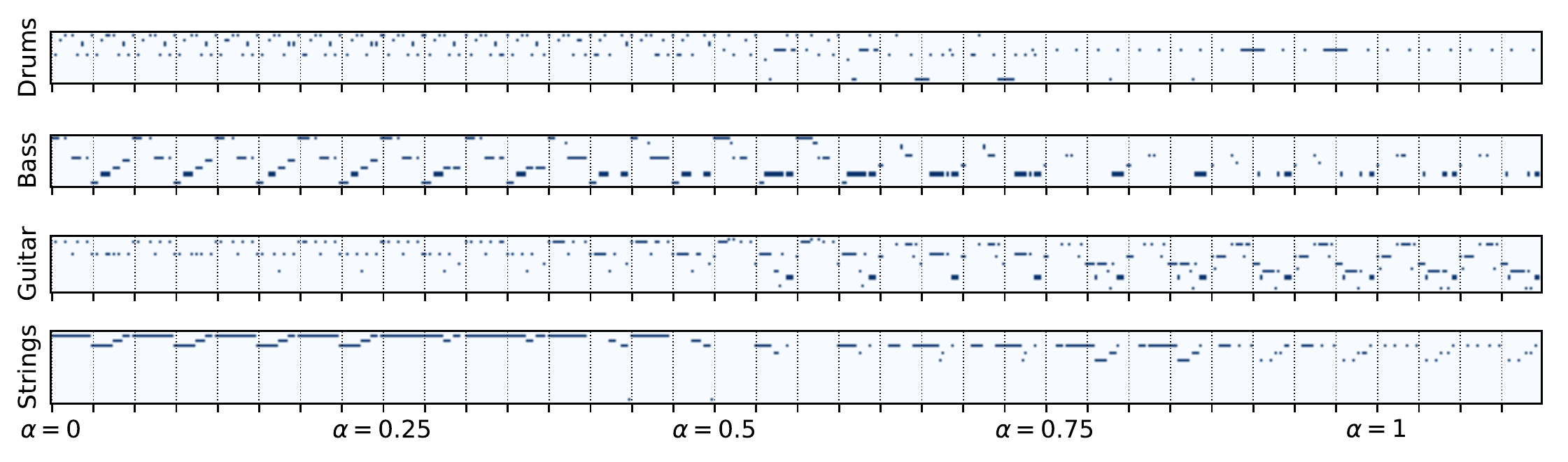}
  \caption{Example of interpolation between two different songs. We notice that the generated pianorolls are gradually changing in a plausible way.}
  \label{fig:interpolation_sample}
\end{figure*}


\subsection{Dataset and Experimental Setup}
\label{rec_exp_setup}
\label{sec_dataset}
We use the same preprocessing procedure of \cite{RobertsEngelRaffel} to perform a direct performance comparison of the reconstruction error with their approach.
The dataset used for the experiments is composed of MIDI songs taken from the ``LMD\_matched" version of the Lakh MIDI dataset \cite{Raffel}, a collection of more than 40.000 song of various genres, transcribed in MIDI format. Each song has a series of metadata associated to it. In our experiments we kept metadata regarding music genres, for an initial total of 1179 different possible genre tags. We then selected the 32 most occurring tags, which accounted alone for more than 35\% of the total genre tags. We use these metadata information to condition the prior distribution of the model's latent space, as described in Sect. \ref{sec_choice_of_prior}.
From this dataset, we retained only songs satisfying the following properties:
\begin{itemize}
\item Songs having only one time signature change event (i.e. songs that did not change their time signature during the execution).
\item Songs having a time signature of $\frac{4}{4}$.
\item Songs containing at least one non-empty drums track (MIDI channel 10).
\item Songs containing at least one non-empty bass track (MIDI program number in the range [32, 39]).
\item Songs containing at least one non-empty guitar/piano track (MIDI program number in the range [0, 31]).
\end{itemize}
A sliding window of size of 2 bars and a stride of 1 bar is used to extract all unique 2-bar (32 timesteps) musical sequences from the remaining songs. If a track has more than one note playing at the same time, we consider only the note with the lowest pitch. This was done for two reasons: first, it let us directly compare quantitative results with similar models found in the literature. Second, finding an effective way to generate completely polyphonic pianorolls is an interesting open problem that deserves a discussion on its own, designing solutions to this is, therefore, of out of the scope of this paper (we refer the interested readers to the hybrid RNN-RBM model of \cite{boulangerlew2012modeling}).
We perform data augmentation on the extracted sequences by randomly transposing every pitch of the sequence to any of the 12 possible tonal keys of the octave, uniformly sampling the amount of transposition semitones from the interval [-5, 6]. If a song contains more than one drums, bass, or guitar/piano tracks, we consider the cross product of all possible combinations. Other tracks that cannot be categorized as either drums, bass or guitar tracks are then merged in an additional ``strings" track. Among the extracted sequences, we only add to the final dataset those containing at most 1 bar of consecutive silence in every track. This preprocessing procedure resulted in about 300,000 data samples.
To test the model's reconstruction ability with very long musical sequences we evaluate on an additional dataset with samples of length 16 bars, corresponding to 256 timesteps. The preprocessing procedure for this dataset is almost identical to the one described above for 2-bar sequences, except for the fact that it uses a 16-bar sliding window with a stride of 1-bar to extract the unique musical sequences from the songs. All other steps are unchanged. This dataset resulted in about 100,000 data samples.
During training, 20\% of samples are retained from their respective datasets as a separate validation set.
All preprocessing is made with the help of \texttt{pypianoroll} \cite{DongHsiaoYang} and \texttt{pretty\_midi} \cite{RaffelEllis} libraries for Python.\\

The experiments use the Adam optimizer \cite{KingmaBa} with a learning rate of $10^{-4}$ and an exponential decay rate of $10^{-4}$, applied after each gradient update. The batch size is 256.
The adversarial regularization weight $\beta$ is annealed from 0 to 1 during training with fixed increases of $0.1$, scheduled every 10,000 gradient updates. The model processes the song in chunks of either 2 or 16 bars (depending on the specific dataset) and each bar is represented by a sequence of 16 timesteps, sampled at regular intervals. During model selection, we found out that the training procedure is very stable and mostly insensitive to the initial random seed or the specific hyperparameter values. The only parameter that required a slightly more careful tuning was the learning rate. In general, we chose the hyperparameter values that resulted in a good convergence speed without adding too much complexity to the training procedure.

We trained one model on the 2-bar dataset, forcing a gaussian prior on the latent space. For the 16-bar dataset, we trained two different models. The first one uses a single gaussian prior distribution, whereas the second one uses the mixture of 32 gaussians conditioned on the genres tag information, described in section \ref{sec_choice_of_prior}. The experiments have been implemented in Python 3.6, using the Keras library (with TensorFlow backend) \cite{keras}\cite{tensorflow}.

\begin{figure*}
\centering
\begin{minipage}{.5\textwidth}
  \centering
  \includegraphics[width=0.9\linewidth]{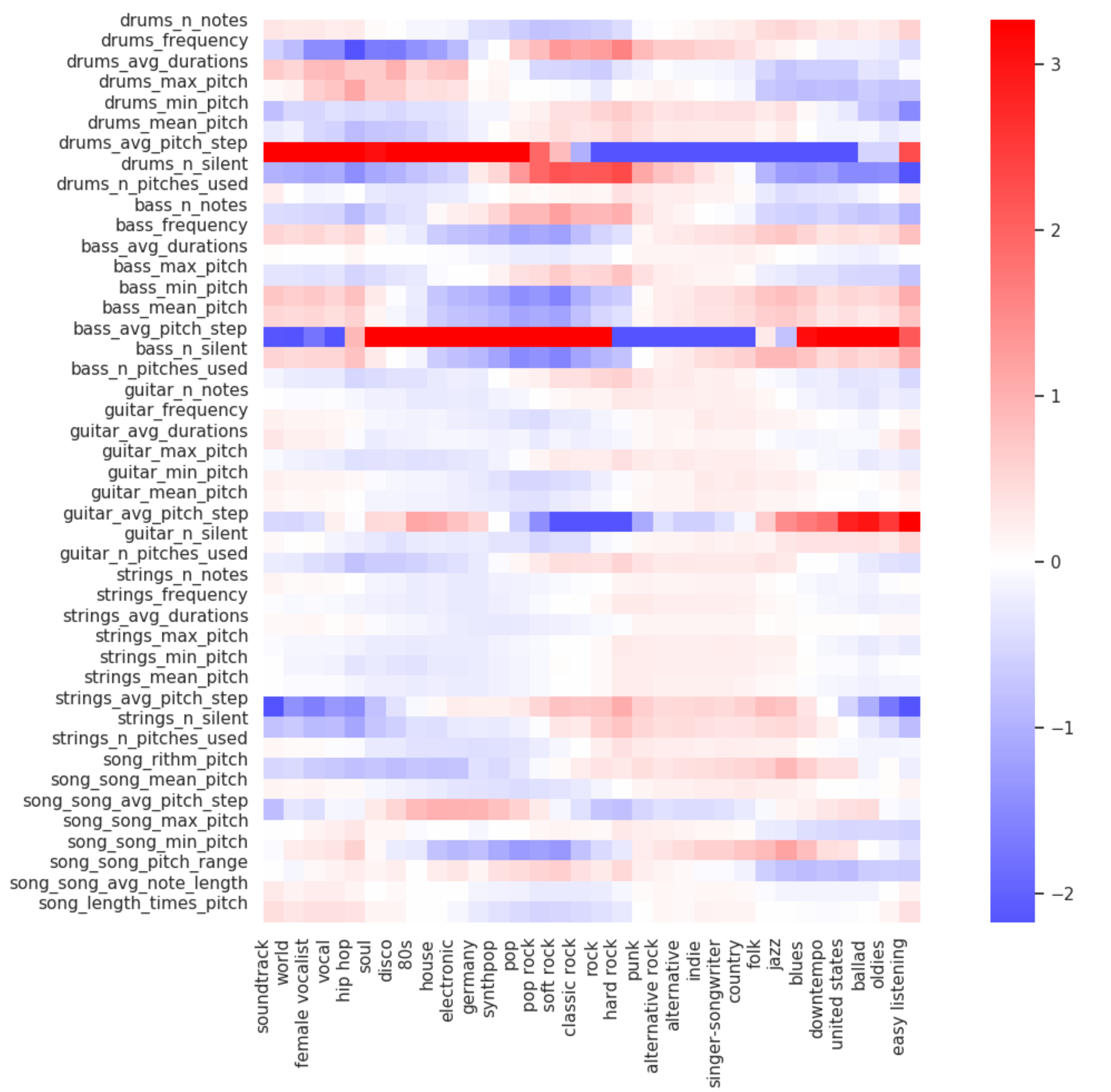}
  \captionof{figure}{Relative differences of low-level metrics of each gaussian in the gaussian mixture prior distribution, computed with respect to their means.}
  \label{fig:relative_changes}
\end{minipage}%
\begin{minipage}{.5\textwidth}
  \centering
  \includegraphics[width=.9\linewidth]{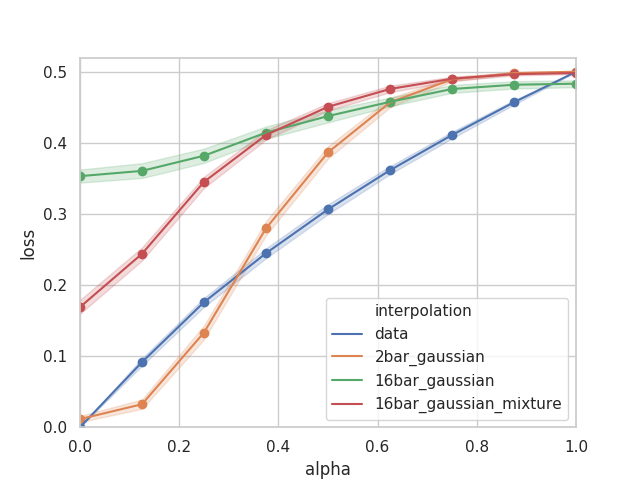}
  \captionof{figure}{Hamming losses of interpolations with respect to $x_1$. The interpolating parameter $\alpha$ is gradually increased from 0 to 1 over time. The result refer to the mean of 256 interpolations.}
  \label{fig:interpolation_distance}
\end{minipage}
\end{figure*}

\subsection{Input Reconstruction}
\label{sec_recon_analysis}

To test the reconstruction abilities of MusAE, a test sample $x$ is first fed as input to the encoder, producing the corresponding latent code $z$, which is then used by the decoder to generate the reconstructed sample $\tilde{x}$. The generated data sample is then compared to the original one. Ideally, if the training was successful, $x$ and $\tilde{x}$ should be indistinguishable from each other.

The final reconstruction accuracy mainly depends on two hyperparameters: the value of the $\beta$ coefficient for the WGAN-GP loss and the number of latent factors. Having high values of $\beta$ ($\beta >> 1$) will cause the model to deviate very little from the imposed prior, using as few latent dimensions as possible and only if it significantly improves reconstruction accuracy. On the other hand, a small $\beta$ allows the model to focus more on reconstruction, at the expense of the entanglement of the latent space \cite{higgins2016beta_vae}\cite{burgess2018understanding_beta_vae}. Similarly, using few latent variables $z$ will force the model to be computationally efficient and to encode in the latent space only the most relevant variation of the data, allowing for better control of generated songs at the expense of the reconstruction accuracy.

Table \ref{tab_accuracy} reports the reconstruction accuracy of the MusAE models over the Lakh MIDI dataset compared to the Music VAE models by \cite{RobertsEngelRaffel}. The gaussian version of MusAE consistently outperforms the similar, flat version of Music VAE, both on the 2-bar and the 16-bar dataset, showing that the AAE framework is a competitive approach for dealing with symbolic music data. On the other hand, to compete with the more complex architecture of hierarchical Music VAE, a more expressive prior distribution is needed. Indeed, the Gaussian mixture version of MusAE reaches a better reconstruction accuracy than the Music VAE. Note that the MusAE models under consideration only differ for the choice of the imposed prior distribution. Thus, this improvement of performances comes at no additional computational cost (except for the increased cost of sampling from a more complicated distribution, which most of the time is negligible). This is another advantage of MusAE over similar models such as Music VAE, where the increased performance is often obtained through a more complex architecture, which necessarily entails a more complex and lengthier training phase.

\subsection{Latent Space Interpolation}
Latent space interpolations allow mixing two different pianorolls. A successful mix performs a gradual change from the initial pianoroll to the final one.
Interpolations are constructed by encoding two selected data samples $x_1$ and $x_2$, thus producing the corresponding factors $z_1$ and $z_2$ in the latent space. The intermediate latent codes are produced using a linear interpolation between the two original latent factors
\begin{equation}
    \hat{z} = \alpha z_1 + (1-\alpha) z_2
\end{equation}
where $\alpha$ is made to vary in the interval $[0, 1]$. The interpolated latent codes $\hat{z}$ are fed to the decoder, which generates the mixed piano rolls.
Figure \ref{fig:interpolation_distance} shows the normalized hamming distances of the MusAE models between the resulting interpolated pianoroll and the initial pianoroll (i.e. $x_1$) for various values of $\alpha$. We also computed, as a baseline, an interpolation in the data space, where we treated the pianorolls as matrices of Bernoulli random variables with parameter $\alpha$. The interpolated pianoroll is then obtained by sampling the random variables. The hamming losses between the interpolated and the final pianoroll $x_2$ behave in a specular manner, so they are not shown.
Figure \ref{fig:interpolation_distance} shows that the hamming distances increase monotonically for all models. This is a sign that the changes in the interpolated pianorolls are smooth, and gradually go from $x_1$ to $x_2$. 
In this case, the best-performing model is the 2bar version with Gaussian prior. The hamming loss of the latent space interpolations closely follows the one of data interpolations, although the former shows a slightly sharper change from the initial and the final piece with respect to the latter. This is aligned with the perception of humans, which usually favour interpolations that combine the two pieces more interestingly than by simply taking an average. The 16bar models have lower reconstruction accuracy than the 2bar model, so they tend to start with a higher hamming distance of the interpolation for $\alpha=0$. Still, the Gaussian mixture prior seems to be beneficial for the interpolation, confirming the results obtained in Section \ref{sec_recon_analysis} when comparing reconstruction errors.
In general, interpolations are smooth and sound quite convincing to the ear \footnote{\url{http://bit.ly/MusAE\_interpolations}}. The interpolated bars are musically meaningful and contain references to both musical pieces. For example, if the starting bar contains few long notes and low pitches, and the ending bar contains many short high-pitched notes, during the interpolation the notes pitch and frequency will gradually increase, while the duration of the notes will gradually decrease. This is a sign that the model successfully learned to map the high-dimensional real manifold into a lower-dimensional, simpler one, where the data samples are evenly distributed in the latent space. An example of interpolation is shown in Figure \ref{fig:interpolation_sample}, which clearly shows a gradual change for increasing values of $\alpha$.

\begin{figure*}
  \centering
  \includegraphics[width=\linewidth]{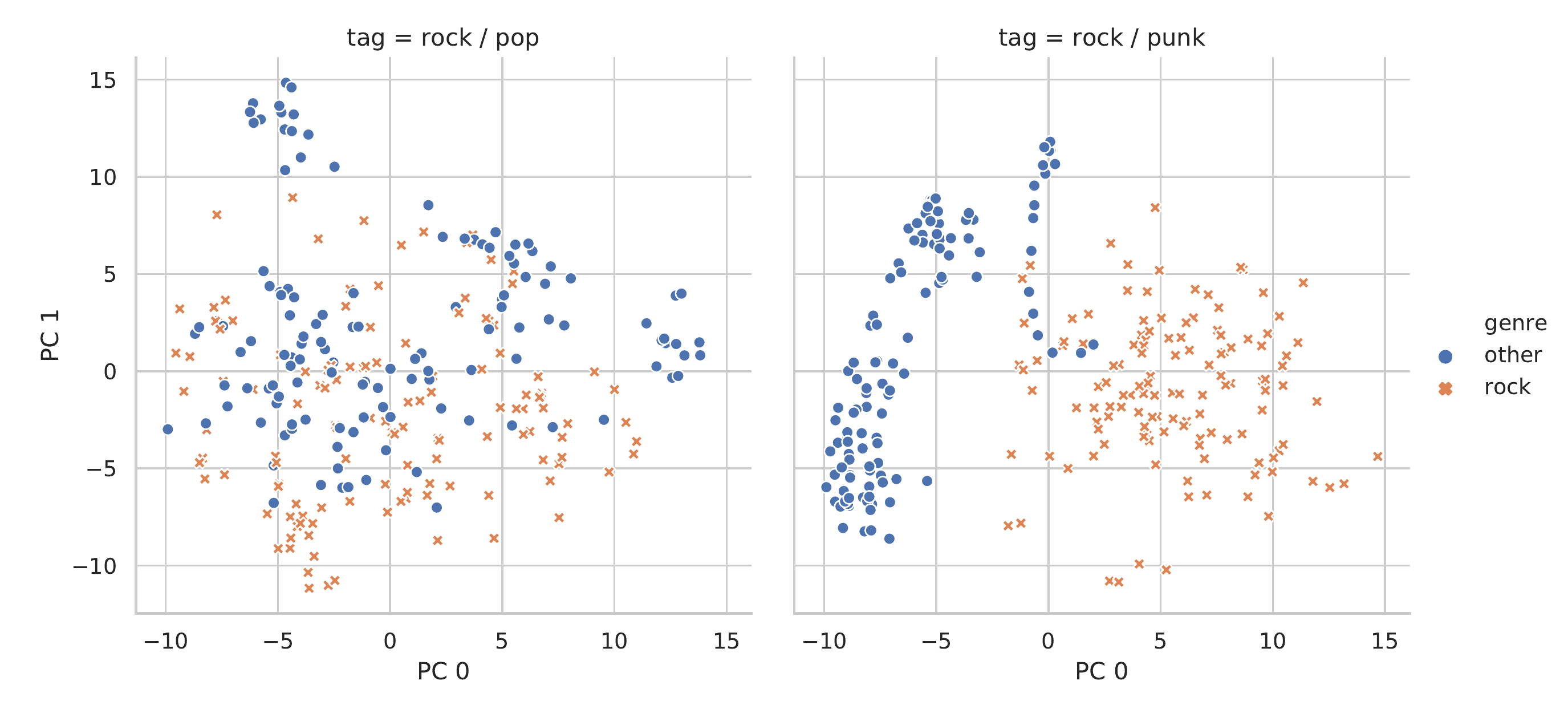}
  \caption{PCA embeddings of pairwise genres.}
  \label{fig:pairwise}
\end{figure*}

\subsection{Latent Space Analysis}
In this experiment, we analyse the behaviour of the latent space when a non-trivial prior distribution is imposed, such as the gaussian mixture distribution previously described in Section \ref{sec_choice_of_prior}. First, we inspect the model's ability of organising its latent space accordingly to low-level characteristics of the pianorolls. Then, we want to see whether it is capable of relating such low-level characteristics to high-level musical concepts, such as genres.

\subsubsection{Low-level Metrics}
We sampled 256 latent vectors from each of the 32 Gaussians that compose the prior distribution. We then decoded the latent vectors and computed a set of metrics over the obtained pianorolls.

These metrics are intended to capture various interesting low-level musical properties, such as the number of notes, average note length, average step between two consecutive pitches, and so on. The complete list of the metrics and their descriptions is reported in the additional material\footnotemark. 
For each metric, we then performed two types of aggregation:
\begin{itemize}
    \item \emph{Global average}: we computed the mean value of the metric over all the pianorolls.
    \item \emph{Local averages}: we computed the mean value of the metric, considering only the pianoroll sampled from a single gaussian. This resulted in 32 mean values, one for each gaussian of the prior distribution.
\end{itemize}
Finally, we computed the relative changes of the local averages with respect to the global average.

The final results are shown in Figure \ref{fig:relative_changes}. The horizontal axis indicates the gaussian of origin of the pianorolls from which the metrics have been computed, while the vertical axis indicates the specific metric. Thus, moving along the horizontal axis of the figure effectively corresponds to moving from a gaussian to the adjacent ones in the latent space (notice that, since in the actual latent space the single gaussian are displaced on a circle, the horizontal axis wraps-up to itself, so that the last gaussian in the figure is adjacent to the first one).
The figure clearly shows a gradual variation of the metrics when moving between adjacent gaussians. This is a further sign that the model successfully learned to organise its latent space in a tidy way: the richer prior distribution allows the model to capture the most peculiar low-level characteristics of each genre and to ``assign" them to the corresponding gaussian. Moreover, this genre-induced organisation of the latent space seems indeed to respect the smoothness property, which, as stated above, is an essential property for being able to support principled and intuitive editing of musical material.

\subsubsection{High-level Genre Information}
Finally, we want to investigate to what extent the genre information, which was injected in an implicit way during the training phase, helped the model to incorporate high-level information about musical style within the latent representations. For each pair of genres, we randomly selected 256 pianorolls from the training set, 128 of which belonging to the first genre and the other 128 belonging to the second genre. Since each song may belong to multiple genres, we decided to select only the pianorolls that are tagged with only one of the two genres considered but not the other. For example, when comparing the ``pop" and ``rock" genre pair, we required the ``rock" pianorolls to not have the ``pop" tag associated with them as well, and vice versa. The selected pianorolls are encoded with the model and projected using the PCA decomposition of the resulting latent codes. The PCA decomposition performs a linear projection along the directions of maximum variance of the latent space of the two genres. Ideally, two different genres should be linearly separable in these reduced space. In this respect, a linear separation would allow the model to perform style transfer more easily. However, most of the musical genres do not have a clear separation, therefore we expect some overlapping between similar genres or between genres that have a very broad definition. Figure \ref{fig:pairwise} shows the first two principal components of two such genre pairs, chosen as examples. The complete plots (and additional plots, obtained using the t-SNE algorithm) are available among the additional material\footnotemark[\value{footnote}]. We denoted the latent codes belonging to the first genre with blue dots, and the ones belonging to the second genre with orange crosses. In general, the results match our intuitive understanding of the relations between the genres. For example, the ``rock vs. pop" plot shows a mixed situation, whereas the ``rock vs. punk" plot presents a clearer separation between the two kinds of pianorolls. This reflects the fact that ``rock" and ``pop" are very broad definitions, which encompasses many musical sub-genres that can be, in practice, quite different from one another. Hence, the dividing line between the two genres is blurry. On the other hand, punk songs tend to present more homogeneous stylistic characteristics, and this results in a clearly distinct ``punk" area inside the ``rock" latent codes. This alignment between latent space projections and human intuition is a hint that the latent space successfully embedded (at least to a certain extent) information about high-level musical style. The fact that even linear techniques such as PCA are already sufficient to visually separate the different genres confirms that, and the global nature of PCA assure us that the separation is real, and not an artifact of the chosen projection method (as it could have happened with t-SNE).r

\footnotetext{\url{http://bit.ly/MusAE\_latent}}

\section{Discussion and Concluding Remarks}
In this paper, we presented MusAE, a novel recurrent AAE that is meant to assist artists by providing a tool for intuitive editing of the musical properties of songs. We showed that an accurate choice of the latent space's prior distribution can considerably improve the model performance, thus avoiding the typical drawbacks entailed by a more complex architecture. Empirical comparisons with similar models show that MusAE has higher reconstruction accuracy of musical sequences than its competitors. Latent space interpolations are smooth and form a convincing transition between the starting and the ending bar. This is a sign that the latent space does not have ``holes", i.e. regions with no direct correspondence in the data space. The latent space seems instead to be evenly populated, and the latent factors are able to generate realistic sequences, independently from the specific part of the latent space they have been sampled from. The experiments showed that the model is indeed able to organise its latent space according to low-level characteristics of the pianorolls, and yet embedding high-level information about musical genres. 

Despite having confirmed our initial hypothesis, this line of research is still at its initial stages and many improvements are possible and worthy of being considered. In the following part of this section, we outline some possible future research directions. While we showed that the latent space is able to take into account genre information when encoding the pianorolls, we still lack a way to leverage such information in a way that allows us to perform effective style transfer between genres in a principled and reliable way. A possible approach for tackling this problem could be trying to find a way to disentangle genre (i.e. style) information from content information in the latent representations. Effective disentanglement of the relevant musical properties can also be beneficial for improving the effectiveness of the model as a smart music editor, for it is usually of interest for the artist to affect only a single specific property, whilst leaving all the other ones mostly untouched (although sometimes a certain degree of entanglement can be useful for maintaining the overall coherence of the musical piece). Finally, the additional flexibility of the AAE regarding the choice of prior distributions can be leveraged even further: for example, it might be useful to have one or more additional sets of latent factor, each one following different prior distributions, to incorporate different relevant metadata information into the latent space. In the end, we believe that by explicitly representing genre information into the latent space, our method can be a first step towards a principled solution to one of the most interesting problems in automatic music generation, that is, genre style transfer.


\bibliographystyle{ecai}
\bibliography{ecai}

\begin{thebibliography}{10}

\bibitem{tensorflow}
Mart\'{\i}n Abadi, Ashish Agarwal, Paul Barham, Eugene Brevdo, Zhifeng Chen,
  Craig Citro, Greg~S. Corrado, Andy Davis, Jeffrey Dean, Matthieu Devin,
  Sanjay Ghemawat, Ian Goodfellow, Andrew Harp, Geoffrey Irving, Michael Isard,
  Yangqing Jia, Rafal Jozefowicz, Lukasz Kaiser, Manjunath Kudlur, Josh
  Levenberg, Dandelion Man\'{e}, Rajat Monga, Sherry Moore, Derek Murray, Chris
  Olah, Mike Schuster, Jonathon Shlens, Benoit Steiner, Ilya Sutskever, Kunal
  Talwar, Paul Tucker, Vincent Vanhoucke, Vijay Vasudevan, Fernanda Vi\'{e}gas,
  Oriol Vinyals, Pete Warden, Martin Wattenberg, Martin Wicke, Yuan Yu, and
  Xiaoqiang Zheng.
\newblock {TensorFlow}: Large-scale machine learning on heterogeneous systems,
  2015.
\newblock \url{https://www.tensorflow.org/}.

\bibitem{ArjovskyBottou}
Martin Arjovsky and Léon Bottou, `Towards principled methods for training
  generative adversarial networks', {\em arXiv:1701.04862v1}, (2017).

\bibitem{Arjovsky}
Martin Arjovsky, Soumith Chintala, and Léon Bottou, `Wasserstein gan', {\em
  arXiv:1701.07875}, (2017).

\bibitem{boulangerlew2012modeling}
Nicolas Boulanger-Lewandowski, Yoshua Bengio, and Pascal Vincent.
\newblock Modeling temporal dependencies in high-dimensional sequences:
  Application to polyphonic music generation and transcription, 2012.

\bibitem{Briot}
Jean-Pierre Briot, Gaetan Hadjeres, and Francois Pachet, `Deep learning
  techniques for music generation—a survey', {\em arXiv:1709.01620}, (2017).

\bibitem{Brunner}
Gino Brunner, Andres Konrad, Yuyi Wang, and Roger Wattenhofer, `Midi-vae:
  Modeling dynamics and instrumentation of music with applications to style
  transfer', in {\em ISMIR}, (2018).

\bibitem{burgess2018understanding_beta_vae}
Christopher~P. Burgess, Irina Higgins, Arka Pal, Loic Matthey, Nick Watters,
  Guillaume Desjardins, and Alexander Lerchner.
\newblock Understanding disentangling in $\beta$-vae, 2018.

\bibitem{keras}
Chollet, François, et~al.
\newblock Keras, 2015.
\newblock \url{https://keras.io}.

\bibitem{Dong}
Hao-Wen Dong, Wen-Yi Hsiao, Li-Chia Yang, and Yi-Hsuan Yang, `Musegan:
  Multitrack sequential generative adversarial networks for symbolic music
  generation and accompaniment', {\em Proceedings of the 32nd AAAI Conference
  on Artificial Intelligence}, (2018).

\bibitem{DongHsiaoYang}
Hao-Wen Dong, Wen-Yi Hsiao, and Yi-Hsuan Yang, `Pypianoroll: Open source python
  package for handling multitrack pianorolls', {\em ISMIR Late-Breaking Demos
  Session}, (2018).

\bibitem{gatys2016image_style_transfer}
Leon~A Gatys, Alexander~S Ecker, and Matthias Bethge, `Image style transfer
  using convolutional neural networks', in {\em Proceedings of the IEEE
  conference on computer vision and pattern recognition}, pp. 2414--2423,
  (2016).

\bibitem{Goodfellow}
Goodfellow, Pouget-Abadie, Mirza, Xu, Warde-Farley, Ozair, Courville, and
  Bengio, `Generative adversarial nets', {\em Advances in Neural Information
  Processing Systems}, (2014).

\bibitem{Gulrajani}
Gulrajani, Ahmed, Arjovsky, Dumoulin, and Courville, `Improved training of
  wasserstein gans', {\em arXiv:1704.00028}, (2017).

\bibitem{hadjeres2016deepbach}
Gaëtan Hadjeres, François Pachet, and Frank Nielsen.
\newblock Deepbach: a steerable model for bach chorales generation, 2016.

\bibitem{higgins2016beta_vae}
Irina Higgins, Loic Matthey, Xavier Glorot, Arka Pal, Benigno Uria, Charles
  Blundell, Shakir Mohamed, and Alexander Lerchner.
\newblock Early visual concept learning with unsupervised deep learning, 2016.

\bibitem{hu2017controlled_text_generation}
Zhiting Hu, Zichao Yang, Xiaodan Liang, Ruslan Salakhutdinov, and Eric~P Xing,
  `Toward controlled generation of text', in {\em Proceedings of the 34th
  International Conference on Machine Learning-Volume 70}, pp. 1587--1596.
  JMLR. org, (2017).

\bibitem{MIDI}
International MIDI Association, {\em General MIDI level 1 specification}, 1991.
\newblock \url{https://www.midi.org/specifications}.

\bibitem{jing2019style_transfer_review}
Yongcheng Jing, Yezhou Yang, Zunlei Feng, Jingwen Ye, Yizhou Yu, and Mingli
  Song, `Neural style transfer: A review', {\em IEEE transactions on
  visualization and computer graphics}, (2019).

\bibitem{KingmaBa}
Diederik~P. Kingma and Jimmy Ba, `Adam: A method for stochastic optimization',
  {\em arXiv:1412.6980}, (2014).

\bibitem{KingmaWelling}
Diederik~P. Kingma and Max Welling, `Auto-encoding variational bayes', {\em
  arXiv:1312.6114v10}, (2014).

\bibitem{li2017demystifying_style_transfer}
Yanghao Li, Naiyan Wang, Jiaying Liu, and Xiaodi Hou, `Demystifying neural
  style transfer', {\em arXiv preprint arXiv:1701.01036}, (2017).

\bibitem{Makhzani}
Alireza Makhzani, Jonathon Shlens, Navdeep Jaitly, and Ian~J. Goodfellow,
  `Adversarial autoencoders', {\em arXiv:1511.05644}, (2015).

\bibitem{Mogren}
Olof Mogren, `C-rnn-gan: continuous recurrent neural networks with adversarial
  training', {\em arXiv:1611.09904}, (2016).

\bibitem{Raffel}
Colin Raffel, `Learning-based methods for comparing sequences, with
  applications to audio-to-midi alignment and matching', {\em PhD thesis,
  Columbia University}, (2016).

\bibitem{RaffelEllis}
Colin Raffel and Daniel P.~W. Ellis, `Intuitive analysis, creation and
  manipulation of midi data with \texttt{pretty\_midi}', {\em ISMIR Late
  Breaking and Demo Papers}, (2014).

\bibitem{RobertsEngelRaffel}
Adam Roberts, Jesse Engel, Colin Raffel, Curtis Hawthorne, and Douglas Eck, `A
  hierarchical latent vector model for learning long-term structure in music',
  {\em arXiv:1803.05428}, (2018).

\bibitem{Schuster}
Schuster and Paliwal, `Bidirectional recurrent neural networks', {\em IEEE
  Transactions on Signal Processing, 45(11):2673–2681}, (1997).

\bibitem{shen2017_parallel_style_text}
Tianxiao Shen, Tao Lei, Regina Barzilay, and Tommi Jaakkola, `Style transfer
  from non-parallel text by cross-alignment', in {\em Advances in neural
  information processing systems}, pp. 6830--6841, (2017).

\bibitem{Yang}
Li-Chia Yang, Szu-Yu Chou, and Yi-Hsuan Yang, `Midinet: A convolutional
  generative adversarial network for symbolic-domain music generation', {\em
  Proc. ISMIR}, (2017).

\bibitem{Yu}
Lantao Yu, Weinan Zhang, Jun Wang, and Yong Yu, `Seqgan: Sequence generative
  adversarial nets with policy gradient', {\em Proceedings of the Thirty-First
  AAAI Conference on Artificial Intelligence, pages 2852–2858, 2017}, (2017).

\end{thebibliography}
\end{document}